\documentclass[aps,prl,twocolumn,amsmath,amssymb,amsfonts,nofootinbib,long,floatfix]{revtex4}
\usepackage{epsfig,latexsym,bm,epstopdf}

\usepackage{color}

\makeatletter
\newcommand{\vast}{\bBigg@{4}}
\newcommand{\Vast}{\bBigg@{5}}
\makeatother

\DeclareMathAlphabet{\mathpzc}{OT1}{pzc}{m}{it}

\allowdisplaybreaks[4]
\usepackage{amsmath}
\usepackage{amssymb}

\usepackage{bm}
\usepackage{arydshln}
\usepackage{diagbox}
\begin{document}

\title{Is Lensing Amplitude Anomaly in the Cosmic Microwave Background the Evidence of Extremely Low Frequency Primordial Gravitational Wave?}

\author{Wenshuai Liu$^{1}$}
\email{674602871@qq.com}
\affiliation{$^1$School of Physics, Henan Normal University, Xinxiang 453007, China}

\date{\today}


\begin{abstract}
Trajectories of photons of cosmic microwave background (CMB) from the surface of last scattering to us could be deflected by extremely low frequency primordial gravitational wave (PGW). With large scale structure (LSS) producing a smoothing of the acoustic peaks in the power spectrum of the CMB anisotropies through weak lensing, the presence of extremely low frequency PGW could enhance the effect of weak lensing on CMB due to the coupling of extremely low frequency PGW and LSS, thus, give rise to much more smoothing of the spectrum. This may be an natural explanation for the lensing amplitude anomaly observed by Planck, meaning that lensing amplitude anomaly may be the evidence of extremely low frequency PGW.
\end{abstract}




\maketitle


\section{Introduction}
Inflation, which gives rise to not only a flat, homogeneous, and isotropic Universe but also seed perturbations growing to create large scale structure in the Universe \citep{8,9,10,11}, predicts the existence of primordial gravitational waves (PGWs) with a nearly scale invariant spectrum \citep{1,2,3,4,5,6,7}. The detection of PGWs would confirm the inflationary scenario and determine the energy scale, thus, it is of great significance to find tools to detect PGWs. The traditional method to detect extremely low frequency PGWs with range of $10^{-18}$Hz-$10^{-16}$Hz is the B-modes of polarization of cosmic microwave background (CMB) \citep{12,13} induced by extremely low frequency PGWs through Thomson scattering. From observation, it shows that dust in our Milky Way can make foreground contamination, leading to challenges with the method of B-modes. Thus, finding an alternative method of detection is worthy of study. When the frequency of PGW is less than $10^{-18}$Hz, the wavelength can be larger than the horizon of the Universe. Here, PGWs with frequency less than $10^{-18}$Hz are also called as extremely low frequency PGWs.

A new method proposed by Liu \cite{43} uses gravitational lens system with a non-aligned source-detector-observer configuration to detect extremely low frequency PGWs, which shows that the time delay in such gravitational lens system with perturbation from extremely low frequency PGWs could deviate from the one deduced from the theoretical model obviously. This means that such gravitational lens system could be used as a possible long-baseline detector of extremely low frequency PGWs. However, the deflector adopted in \cite{43} is a point mass model which is unrealistic in strong lensing. Work of Liu \cite{46} adopts gravitational lens system with singular isothermal sphere deflector. It shows from Liu \cite{46} that, under the perturbation of extremely low frequency PGWs with arbitrary direction of propagation, the time delay from observation can strongly deviate from that deduced from the theory.

It shows from Liu \cite{46} that extremely low frequency PGWs can affect gravitational lensing through perturbation on the trajectory of photon no matter what the gravitational lensing is strong lening or weak lensing, meaning that extremely low frequency PGWs could also have effect on the weak lensing of CMB through the coupling of extremely low frequency PGWs and LSS.

The CMB radiation undergoes weak lensing shown in Figure 1 by LSS along path since last scattering surface and the effect of weak lensing on CMB is a remapping of the CMB temperature anisotropies and the CMB polarization field, leading to a smoothing of the acoustic peaks in the power spectrum of the CMB anisotropies \cite{47}. With the presence of extremely low frequency PGWs, the effect of weak lensing on CMB may be enhanced by the coupling of extremely low frequency PGW and LSS, thus, it may give rise to much more smoothing of the spectrum. Results from Planck \cite{48} show that the lensing-reconstruction power spectrum shows to be almost consistent with that expected for $\Lambda$CDM models which fit the CMB spectra. However, the amount of smoothing observed in the CMB angular power spectrum is larger than the amount of the smoothing derived from the power spectrum of the reconstructed lensing potential, resulting the lensing parameter $A_L>1$. The observed $A_L>1$ or the observed smoothing which is larger than expected may be the evidence of extremely low frequency PGWs.

In addition to the lensing amplitude anomaly observed in CMB, other anomalies like Hubble tension, $\Omega_k\ne0$ and $S_8$ tension may be also the evidence of extremely low frequency PGWs since $H_0$, $\Omega_K$ and $S_8$ determined based on CMB can be affected by the weak lensing through the coupling of extremely low frequency PGWs and LSS.

\section{Weak lensing by the coupling of PGWs and LSS}
After inflation, PGWs can be stretched to scale larger than the horizon, this makes PGWs with wavelength larger than the horizon not behave like waves but have an almost frozen amplitude, meaning that such PGWs like a constant anisotropy or shear. Such PGWs start oscillating inside the horizon after they re-enter the horizon, and the amplitude are damped due to expansion of the universe.

In order to only investigate the effect of extremely low frequency PGWs on the photon of CMB from the last scattering surface, we first give the metric of extremely low frequency PGWs as
\begin{equation}
h_{ij}=\frac{a_0}{a}[(u_i u_j-v_i v_j)h_++(u_i v_j+v_i u_j)h_\times]\times\cos(\omega \eta-\mathbf{k} \cdot \mathbf{x})\label{1}
\end{equation}
where $a$ is the scale factor and the present value is set to be $a_0=1$. The speed of light is set to be $c=1$. The conformal time $\eta=t_e+(z+L)$ in order to approach the level of approximation. $z$ is in form of conformal distance and conformal time. $L=14000$Mpc is the conformal distance between the last scattering surface and us. $t_e$ is the time when the photon was emitted at $(x=0,y=0,z=-L)$ so that $\omega t_e$ acts as the initial phase, $\mathbf{k}=\omega(\sin\theta\cos\phi, \sin\theta\sin\phi, \cos\theta)$ is the propagation vector, $\mathbf{u}=(\sin\phi, -\cos\phi, 0)$, $\mathbf{v}=(\cos\theta\cos\phi, \cos\theta\sin\phi, -\sin\theta)$, $\omega=2\pi f$, $f$ is the frequency of gravitational wave at present, $h_+$ and $h_\times$ are the amplitude of the two polarizations of the gravitational wave at present, respectively.

In Eq. (\ref{1}), when the wavelength of PGW is larger than the horizon of the universe, it is frozen and does't evolve, resulting $\eta=t_e$ in Eq. (\ref{1}). Only when the wavelength of PGW is smaller than the horizon of the universe can $\eta=t_e+(z+L)$.

\begin{figure}
      \includegraphics[width=\columnwidth]{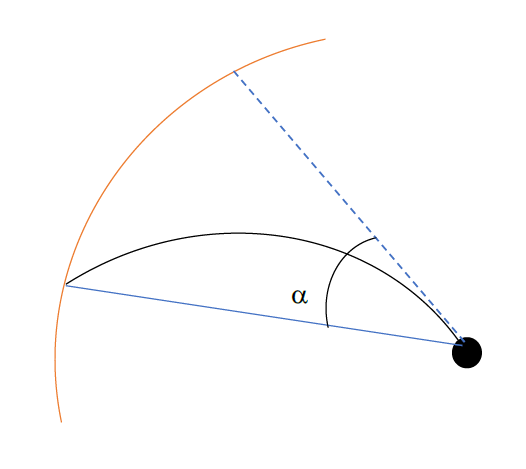}
	\caption{The orange curve represents the last scattering surface and the black dot is the observer. The blue line shows the trajectory of CMB photon without perturbation of LSS and PGWs, and the black curve is the trajectory of CMB photon with perturbation of LSS and PGWs. The blue dashed line is the direction of CMB photon the observer sees.}
    \label{figure1}
\end{figure}

\begin{figure}
      \includegraphics[width=\columnwidth]{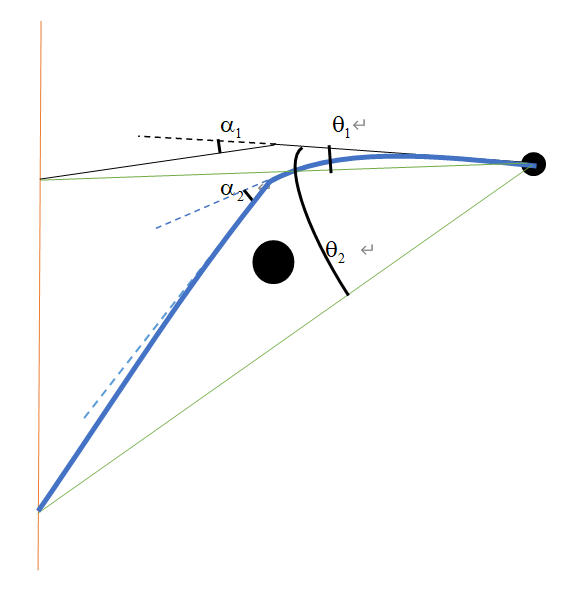}
	\caption{The orange line represents a small fraction of the last scattering surface, the big black dot is a cluster(In reality, there are many clusters between the the last scattering surface and the observer along line of sight. Here, we set one cluster in order to describe the effect of PGWs on lensing) and the small dot is the observer. The black line represents the trajectory of CMB photon with perturbation of the cluster but without perturbation of PGWs. The blue curve is the trajectory of CMB photon with perturbation of the cluster and PGWs. Due to the fact that PGW could deflect the trajectory of the photon, the closest distance between the photon and the cluster may be smaller with perturbation of PGWs compared with that without perturbation of PGWs, thus, the deflection by the cluster can be larger ($\alpha_2>\alpha_1$), resulting $\theta_2>\theta_1$.}
    \label{figure2}
\end{figure}

To show the deflection of the photon of CMB due to extremely low frequency PGWs, the method to calculate the deflection angle is similar to the method of Liu \cite{46} to calculate the angular position of the image of the deflector in the gravitational lens system. When we consider PGW with wavelength larger than (and almost equal to) the horizon of the universe, the frequencies we adopt are $f=10^{-19}$, $f=5\times10^{-19}$, and $f=10^{-18}$, and we set the tensor to scalar ratio to be $r=0.01$, $r=0.005$, $0.001$ to calculate the root-mean-square of deflection angles. For $r=0.01$, the root-mean-square of deflection angles are $6.93423\times10^{-7}$, $3.38361\times10^{-6}$ and $6.15404\times10^{-6}$ when $f=10^{-19}$, $f=5\times10^{-19}$, and $f=10^{-18}$. For $r=0.005$, they are $4.90326\times10^{-7}$, $2.39258\times10^{-6}$ and $4.35158\times10^{-6}$ when $f=10^{-19}$, $f=5\times10^{-19}$, and $f=10^{-18}$. For $0.001$, they are $2.19273\times10^{-7}$, $1.06996\times10^{-6}$ and $1.946023\times10^{-6}$ when $f=10^{-19}$, $f=5\times10^{-19}$, and $f=10^{-18}$.

According to the above, it shows that extremely low frequency PGWs could strongly deflect the photon of CMB. Although the deflection of CMB due to LSS is about 2 arcminutes which is larger than the above, the coupling of extremely low frequency PGW and LSS may enhance the weak lensing shown in Figure 2 and the deflection induced by the coupling of extremely low frequency PGW and LSS could be larger than linear superposition of the deflection induced by LSS and that induced by extremely low frequency PGWs, resulting the observed smoothing of the spectrum may be larger than that derived from theory.

We only give a qualitative analysis of the effect of extremely low frequency PGWs on weak lensing, the detailed quantitative calculation should be conducted to confirm if extremely low frequency PGWs could explain the lensing amplitude anomaly. The method of calculation is based on Boltzmann equation for photon, which is shown as
\begin{equation}
\frac{\partial f}{\partial t}+\frac{\partial f}{\partial x^i}\frac{dx^i}{dt}+\frac{\partial f}{\partial p}\frac{dp}{dt}+\frac{\partial f}{\partial n^i}\frac{dn^i}{dt}=C(f) \label{2}
\end{equation}

In Eq. (\ref{2}), the term $\frac{\partial f}{\partial n^i}\frac{dn^i}{dt}$ leads to weak lensing. $\frac{dn^i}{dt}$ can be derived based on the geodesic equation
\begin{equation}
\frac{d^2x^\mu}{d\lambda^2}+\Gamma^\mu_{\alpha\beta}\frac{dx^\alpha}{d\lambda}\frac{dx^\beta}{d\lambda}=0
\end{equation}
where
\begin{equation}
\Gamma^\mu_{\alpha\beta}=\frac{1}{2}g^{\mu\rho}(\partial_\alpha g_{\rho\beta}+\partial_\beta g_{\rho\alpha}-\partial_\rho g_{\alpha\beta})
\end{equation}
and $g_{\mu\nu}$ can be obtained from
\begin{eqnarray}
ds^2=a^2(\eta)[-(1+2U)d\eta^2+(1-2U)(dx^2+dy^2+dz^2)
\nonumber \\
+h_{ij}dx^idx^j]
\end{eqnarray}
Then, with code CAMB \cite{49} and CLASS \cite{50,51} modified to include the effect of weak lensing from the coupling of extremely low frequency PGWs and LSS, the detailed confirmation could be as follows.

Using the coupling of density perturbation derived from $\Lambda$CDM and tensor perturbation determined by scalar to tensor ratio $r$ to numerically generate the CMB map and calculate the CMB temperature power spectrum, we can infer the scalar to tensor ratio $r$ by setting the calculated CMB temperature power spectrum to be consistent with the observed temperature power spectrum. If the inferred $r$ is larger than the observed value, the lensing amplitude anomaly may be not due to PGWs. If the derived $r$ is smaller than the observed value, the lensing potential should be reconstructed based on the generated CMB map in order to get the CMB temperature power spectrum and the amount of smoothing based on the power spectrum of the reconstructed lensing potential. If the resulting amount of smoothing is consistent with the amount of smoothing derived from the power spectrum of the reconstructed lensing potential in Planck \cite{48} and if the power spectrum of the reconstructed lensing potential based on the generated CMB map is consistent with that reconstructed in Planck \cite{48}, it means that extremely low frequency PGWs may account for the lensing amplitude anomaly.


\section{DISCUSSION} \label{discuss}
The effect of extremely low frequency PGWs on weak lensing of CMB by LSS is discussed qualitatively. It shows that the effect of weak lensing on CMB may be enhanced by the coupling of extremely low frequency PGWs and LSS if extremely low frequency PGWs exist. Since weak lensing by LSS can lead to a smoothing of the acoustic peaks in the angular power spectrum of the CMB anisotropies, thus, weak lensing by the coupling of extremely low frequency PGWs and LSS may give rise to much more smoothing of the spectrum.

Observations by Planck \cite{48} show many anomalies such as lensing amplitude anomaly observed in CMB, Hubble tension, $\Omega_K\ne0$ and $S_8$ tension. From the point of view of weak lensing of CMB, the presence of extremely low frequency PGWs may account for the lensing amplitude anomaly since the coupling of extremely low frequency PGWs and LSS could enhance the effect of weak lensing of CMB. Due to the fact that $H_0$, $\Omega_K$ and $S_8$ determined based on CMB can be affected by the coupling of extremely low frequency PGWs and LSS, anomalies like Hubble tension, $\Omega_K\ne0$ and $S_8$ tension may also reflect the existence of extremely low frequency PGWs.

In this work, we give a qualitative investigation on how the coupling of extremely low frequency PGWs and LSS may enhance the effect of weak lensing on CMB and wonder if the lensing amplitude anomaly might be the evidence of extremely low frequency PGWs. A detailed quantitative verification on extremely low frequency PGWs accounting for lensing amplitude anomaly and other anomalies like Hubble tension, $\Omega_K\ne0$ and $S_8$ tension is in a companion paper which will appear soon in the near future.


\begin{thebibliography}{99}

\bibitem[Grishchuk(1976)]{8}Grishchuk, L. P.\ 1976, JETP Lett. 23, 293

\bibitem[Grishchuk(1977)]{9}Grishchuk, L. P.\ 1977, Sov. Phys. Usp. 20, 319

\bibitem[Starobinsky(1980)]{10}Starobinsky, A. A.\ 1980, PLB 91, 99


\bibitem[Linde(1982)]{11}Linde, A. D.\ 1982, PLB 108, 389

\bibitem[Abbott \& Wise(1984)]{1}Abbott, L. F., \& Wise, M. B.\ 1984, Nucl. Phys. B244, 541
\bibitem[Starobinskii(1985)]{2}Starobinskii, A.\ 1985, Sov. Astron. Lett. 11, 133
\bibitem[Rubakov et al.(1982)]{3}Rubakov, V. A., Sazhin, M.V., \& Veryaskin, A.V.\ 1982, Phys. Lett. 115B, 189

\bibitem[Fabbri \& Pollock(1983)]{4}Fabbri, R., \& Pollock, M. D.\ 1983, Phys. Lett. 125B, 445
\bibitem[Starobinsky(1979)]{5}Starobinsky, A. A.\ 1979, JETP Lett. 30, 682
\bibitem[Sahni(1990)]{6}Sahni, V.\ 1990, \prd 42, 453
\bibitem[Allen(1988)]{7}Allen, B.\ 1988, \prd 37, 2078










\bibitem[Kamionkowski et al.(1997)]{12}Kamionkowski, M., Kosowsky, A., \& Stebbins, A.\ 1997, \prl 78, 2058

\bibitem[Seljak \& Zaldarriaga(1997)]{13}Seljak, U., \& Zaldarriaga, M.\ 1997, \prl 78, 2054


\bibitem[Liu(2022)]{43}Liu, W.,\ 2022, MNRAS, 517, 2769
\bibitem[Liu(2023)]{46}Liu, W.,\ 2023, arXiv:2212.02221

\bibitem[Seljak(1996)]{47}Seljak, U.,\ 1996, ApJ, 463, 1


\bibitem[Lewis et al.(2000)]{49}Lewis, A., Challinor, A. \& Lasenby, A.\ 2000, \apj 538,473
\bibitem[Lesgourgues(2011)]{50}Lesgourgues, J.\ 2011, arXiv:1104.2932
\bibitem[Blas et al.(2011)]{51}Blas, D., Lesgourgues, J., \& Tram, T.\ 2011, arXiv:1104.2933


\bibitem[Aghanim et al.(2020)]{48}Aghanim, N., Planck Collaboration VI.\ 2020, A\&A, 641, A6









\end{thebibliography}
\end{document}